\documentclass[runningheads]{llncs}
\usepackage{graphicx}
\usepackage{makecell}
\usepackage{pdflscape,array,booktabs}
\usepackage{enumitem}
\usepackage{outlines}
\usepackage{tabularx}
\usepackage{longtable}
\usepackage{mdframed}
\usepackage{soul}
\usepackage{xurl}
\usepackage{multirow}
\usepackage{subcaption}
\usepackage[textsize=scriptsize,backgroundcolor=yellow!40]{todonotes}

\begin{document}
\title{Distributed Trust Through the Lens of Software Architecture}
%
%
\author{Sin Kit Lo
\and
Yue Liu
\and
Guangsheng Yu
\and
Qinghua Lu
\and
Xiwei Xu
\and Liming Zhu}
\authorrunning{SK. Lo et al.}
%
\institute{Data61, CSIRO, Sydney, Australia}
\maketitle              
\begin{abstract}
Distributed trust is a nebulous concept that has evolved from different perspectives in recent years. While one can attribute its current prominence to blockchain and cryptocurrency, the distributed trust concept has been cultivating progress in federated learning, trustworthy and responsible AI in an ecosystem setting, data sharing, privacy issues across organizational boundaries, and zero trust cybersecurity. This paper will survey the concept of distributed trust in multiple disciplines. It will take a system/software architecture point of view to look at trust redistribution/shift and the associated tradeoffs in systems and applications enabled by distributed trust technologies.

\keywords{Distributed trust \and Responsible AI \and Software architecture \and Federated learning \and Blockchain \and Pattern \and Machine learning \and Artificial Intelligence.}

\end{abstract}
\section{Introduction}
In recent years, large-scale computation has been essential for data management and processing, and thanks to the advancement in technology, all these can be achieved through multi-party processes \cite{van2016}. The wide adoption of distributed system design to build real-world web applications triggers the consideration of digital trust across multiple nodes while involving multiple stakeholders. Furthermore, the recent emergence of various data-sharing mechanisms such as decentralized, distributed and federated settings (e.g., blockchain and federated learning) further increase the system's complexity. These multiple yet distributed entities, each running different processes need to collect, share, and utilize the data may invoke the privacy, security, and trust risks of adversarial parties compromising the entire system, data, and ultimately the stakeholders. Hence, trust and trustworthiness of distributed systems have generated much attention from academia and industries, harnessing new concepts such as trustworthy and responsible AI, and zero-trust architecture \cite{mckinsey22}. For instance, IBM and Microsoft are setting up a new partnership to meet cybersecurity challenges in the hybrid multi-cloud, specifically using a zero-trust methodology \cite{mccurdy_2021}. All these have contributed to the effort of realizing distributed trust.

\subsection{Distributed Trust}

Digital trust has been discussed across many works of literature which led to various definitions and frameworks available to define and examine the concept. Trust is often conceptual, being used frequently and intuitively yet remains intriguing to define and study. For instance, the World Economic Forum defined digital trust as ``individuals'' expectation that digital technologies and services, and the organizations providing them, will protect all stakeholders’ interests and uphold societal expectations and values'' \cite{initiatives.weforum.org}. McKinsey mentioned in their ``McKinsey Technology Trends Outlook 2022 \cite{mckinsey22}'' that digital trust addresses digital risk across data, cloud, AI and analytics, and risk culture while digital trust technologies empower organizations to gain advantages by building, scaling, and maintaining the trust of stakeholders while using their data, and digital-enabled services. Ozkaya has expressed trust as a crosscutting software concern, where building trust in software systems requires software engineers to consider how they are built by understating different trust-related attributes; organizational practices that are applied in the development, operations, and sustainment of systems; users’ perception of these systems; and related data and its management \cite{9758585}. According to Siau and Wang, trust is defined as (1) being a set of specific beliefs dealing with benevolence, competence, integrity, and predictability (trusting beliefs). (2) the willingness of one party to depend on another in a risky situation (trusting intention); or (3) the combination of these elements \cite{siau2018building}. Thus, we can view trust as a metric to measure how trustworthy a digital or software system is. It is worth noting that trust and trustworthiness are distinct from one another but closely related. Essentially, the difference is between what a system can objectively perform (trustworthiness) and what a truster/stakeholder “prefers/wants” (trust expectation) and their subjective estimation of the behavior of the AI system \cite{Zhu2022}. Trust is about subjective estimation and perception and is not limited to the system’s trustworthiness properties. In this work, we will mainly focus on trustworthiness principles and not the narrow definition of trust (subjective expectation/estimation). 

With the increase in system scale and complexity, centralized development and management encounter more efficiency and reliability challenges. Furthermore, the increase in data and data providers encourages the utilization of distributed system design. However, the distribution also introduces more stakeholders to the system and the complexity of authorities and institutional power also increases where trustworthiness between the stakeholders and the management of authority becomes crucial to enable system trust. As the system settings transition to more decentralized manners,  more dilution of institutional authority from the centralized entities is practiced, and hence, the trust is further distributed to individual stakeholders of the system.  

Ultimately, distributed trust examines the fulfillment of different trustworthiness principles, attributes, and qualities by a software system while taking into consideration the centralized, distributed, or decentralized nature of stakeholders, software layers, components, and data that interact with one another for the operation of the system.    

\begin{figure*}[h!]
\begin{center}
\footnotesize
\begin{mdframed}[
    skipabove=1cm,    
    innerleftmargin = 0.2cm,
    innerrightmargin= 0.2cm,
    usetwoside=false,
]

\begin{quotation}

\textit{\textbf{Definition of Distributed Trust:}} The achievement of expectations that digital technologies and services will protect all stakeholders’ interests and uphold societal values through distributed and decentralised stakeholder authorities and systems.
\end{quotation}

\end{mdframed}
\end{center} 
\end{figure*}

\subsection{Centralized vs Distributed vs Decentralized Systems}

To better understand the trustworthiness and trust of current software systems, we need to clarify the different system settings and the challenges in each of them to achieve trust. The allocation of responsibility consists of two main types: (1) the centralization/distribution of data or/and computations, and (2) the centralization/decentralization of authorities. Conventionally, centralized computing systems have all the tasks, computation, authorities, and data under the same entity. The ease of system development and management of centralized systems comes with low system efficiency. Hence, distributed computing settings are introduced. Here, However, the authority of a distributed system may still be centralized. Some examples of such distributed systems are distributed computing/systems; federated learning. Distributed computing systems are computer systems that consist of multiple interconnected devices that work together to perform a common task while still being managed by a single entity. These distributed devices can be located in the same physical location or in different locations and they communicate with each other over a network to perform the task \cite{van2002distributed}. A distributed computing system can be used to improve the performance and availability of a system, by allocating the workload among multiple devices. One of the key features of distributed computing systems is their ability to process large amounts of data in parallel. This is achieved by dividing the data into smaller chunks and distributing them among the different devices in the system. The devices then work on their assigned chunks of data simultaneously and return their results to a central location. This allows for much faster processing times than possible with a single device \cite{dean2008mapreduce}. Another important feature of distributed computing systems is their ability to provide high availability and fault tolerance. This is achieved by replicating data and processing across multiple devices so that if one device fails, the system can continue to operate using the remaining devices. This ensures that the system is able to continue to operate even in the event of a failure, reducing downtime and improving reliability.

While distributed systems have higher efficiency compared to centralized systems, the centralized authority and management nature have triggered trustworthiness, reliability, and transparency concerns. Considering the decentralization of the authority and management of a system to enable trust, decentralized systems are introduced. If the trust of the centralized systems is considered local by a single stakeholder, the trust of the distributed systems needs to be considered at an institutional and organizational level, while decentralized systems further distribute the institutional power to what we define as distributed trust.

Decentralized computing systems are systems in which multiple independent devices work together to perform a common task, having identical capabilities and responsibilities and all communication is symmetric \cite{10.1007/3-540-45518-3_18}. However, one key difference between conventional distributed systems and decentralized systems is that decentralized systems do not rely on a central authority or single point of failure and instead rely on the coordination of the devices to achieve a common goal. There are two major types of decentralized allocation of responsibility: (1) decentralized authority of the system and distribution of data/compute, such as distributed ledger/smart contract/DAO. (e.g., proof-of-work/stake consensus algorithm) and (2) decentralized authority, centralized data/compute, such as multi-org/institutions having a governance process to decide on how centralized data/compute is done (e.g. time-share/resource allocation in a High-Performance Computer/Mainframe). Decentralized computing systems use consensus algorithms, which are used to ensure that all devices in the system are in agreement about the state of the system \cite{8029379,5072249}. Examples of consensus algorithms include proof-of-work and proof-of-stake \cite{7930224}. These algorithms allow devices in the system to reach an agreement on the state of the system without the need for a central authority. A classic example of a decentralized computing system is a peer-to-peer (P2P) network, in which multiple devices can share resources and communicate without needing a central server. The P2P connection topology is widely adopted in realizing public blockchain-based systems~\cite{LI2018133,7930224}. In a P2P infrastructure, the traditional distinction between clients and back-end servers is simply disappearing \cite{10.1145/502585.502638}. Without a central server as the system coordinator for all the participating agents, mutual trust between the agents becomes the key to their cooperation in a game-theoretic situation which increases the aggregated utility for the participating agents~\cite{10.1145/502585.502638}. Each agent has to obtain information from other peers and propagate the information to other agents through neighboring peers. Hence, it is significant to identify if each agent has reliable peers. In reality, each peer might be faulty or might send obsolete, incorrect, or adversarial information to the other peers, ultimately resulting wrong decision~\cite{Spaho2914,Aikebaier2011}.

From a trust vs ease of management perspective, if the centralized authority is trustworthy or has fulfilled certain trustworthiness principles (e.g., transparency, etc), we can consider centralized authority trustworthy and efficient to manage. However, the current trend of the world is moving away from trusting any centralized authority and needing to meet diverse “stakeholder” expectations so decentralized authority is equally important. Any decentralization of authority may introduce complexity but consensus algorithms are able to solve that complexity by automating the process to some degree. However, to enable consensus algorithms, the blockchain governance patterns show another layer of coordination above the consensus algorithm is needed. For instance, which algorithm/code version to use, who decides, who has commit privilege, and how to define/pick “X” in proof-of-X. Distribution of data/compute is largely limited by computing/data constraints. Data also cannot be moved to a central place or controlled by different entities. Hence, managing a distributed system also introduces additional complexity of cost.

\begin{figure}
    \centering
    \includegraphics[width=0.8\linewidth]{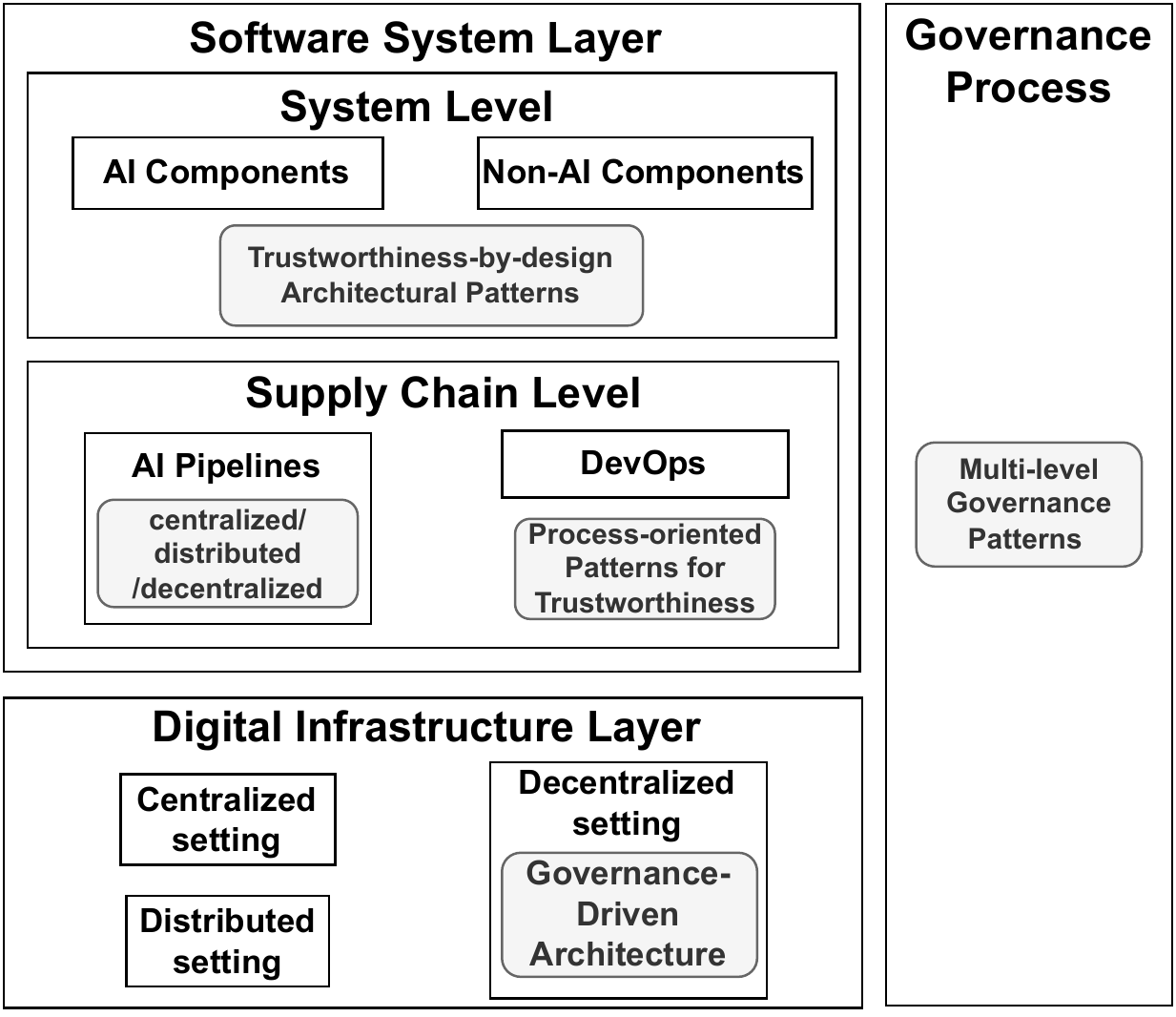}
    \caption{Distributed Trust Architecture Layers.
    \label{Fig:TRUSTARCH}}
\end{figure}

Table~\ref{tab:comparisons_trust} briefly summarized the authority and orchestration, ease of management, and coverage of trust for centralized, distributed, and decentralized systems. Fig. \ref{Fig:TRUSTARCH} showed a multi-layer, distributed trust architecture. The distributed trust architecture represents the different types of system infrastructure settings, covering centralized, distributed, and decentralized settings. On top of the digital infrastructure layer is the software system layer. Under the software system layer, we break down further into two level: (1) system level that handles the AI/ML training pipelines, overall DevOps components such as data, code, and model version control, and any other third-party components for the operation of the system; and (2) supply chain level that consists of AI components (e.g., centralized, distributed, and federated learner) and non-AI components (e.g., data processor, inference APIs, etc.). Overall, each layer is governed by multi-level governance processes to enable distributed trust. Patterns can also be embedded and applied to individual components or layers to enable trust through system-level solutions. For instance, we embedded \textit{trustworthiness-by-design architectural patterns} in the system level under the software system layer to enable trustworthiness-by-design solutions. We also embedded \textit{process-oriented patterns for trustworthiness} to monitor and manage the compliance of the supply chain level components towards achieving distributed trust. The detailed elaboration of these pattern groups will be covered in the next section.

\begin{table}[t!]
\caption{Trust of Centralized vs Distributed and Decentralized Systems}
\begin{tabular}{lccc}
{} & {\footnotesize \textbf{Centralized}} & {\footnotesize \textbf{Distributed}} & {\footnotesize \textbf{Decentralized}}\\
\hline
\footnotesize
\textbf{Authority} & \footnotesize Single-party & \footnotesize Single or multi-party & \footnotesize Multi-party \\
\footnotesize
\textbf{Ease of management} &  \footnotesize High & \footnotesize Moderate & \footnotesize Low \\
\footnotesize
\textbf{Trust Coverage} & \footnotesize Local & \footnotesize Institutional & \footnotesize Distributed \\

\label{tab:comparisons_trust}
\end{tabular}
\end{table}

This paper intends to survey and review the concept of distributed trust in multiple disciplines. We consider mainly the architecture point of view to look at trust redistribution/shift and the associated tradeoffs in systems and applications enabled by distributed trust technologies. The contributions of this paper are:

\begin{itemize}
    \item A summary of trust principles defined by multiple renowned organizations, specifically covering the different trust challenges in distributed trust for centralized, distributed, and decentralized AI and data transaction systems. 
    
    \item A distributed trust architecture as a basis to review the various state-of-the-art system-level solutions (design patterns/tactics, reference architecture, etc.) and process-level solutions (governance, policies, assessments, etc.) to address various challenges in realizing distributed trust for AI-based and data transaction systems.

\end{itemize}

\section{Achieving Distributed Trust}

Considering the requirements and concerns towards distributed trust, we need to understand how to achieve it, with a subjective perception imposed by one key constraint whereby the system is distributed, via trustworthy ways and its challenges. We classify the methods to achieve distributed trust into two main categories: (1) achieving trustworthiness and trust from the product (architecture/design patterns) and achieving trustworthiness principle realization through governance process (institutional governance mechanisms and developing processes/patterns), such as ethical AI, responsible AI or trustworthy AI principles for development or governance guidelines. Despite being two distinct categories of methods to achieve trust, they are mostly intertwined and complement one another in achieving distributed trust.

\subsection{Achieving Distributed Trust Through Product}

Software products are developed as solutions to fulfill various functional and non-functional requirements of a software system. One of them is architecture or design patterns. Patterns are sets of reusable solutions documented to a commonly occurring issue under a given context during software architecture design \cite{bass2003software,lo2021architectural,10.1007/978-3-030-86044-8_6,9426788}. They are usually composed of one or more architecture tactics, which is an implementable architecture design decision to fulfill certain requirements \cite{bass2003software}. Patterns have been widely adopted by software architects and researchers to collect and summarise the existing best practices built on tested software as guidance or reference to build a software architecture or system. With evaluations of consequences and implementation tradeoffs, patterns clearly state the reasons why specific design choices were made. Therefore, patterns can help to resolve this conflict by allowing developers to focus on key design decisions and their decision rationale. 

Another software product is the software architecture design. Software architecture is a step in the Software Development Life Cycle (SDLC) that defines ways to tackle design problems, fulfilling the functional and non-functional requirements \cite{10.1007/978-3-540-24769-2_14,SHARMA201516}. To effectively design comprehensive software architectures, software architects often refer to readily available reference architectures of similar or related software systems. Reference architecture is a standard guideline for system designers and developers for the selection of best practice solutions during the design of software architecture for their problems, which can be further customized as required \cite{10.1007/978-3-030-86044-8_6}.

To enable distributed trust through software products, the elements of trust or a system as the system's non-functional requirements need to be fulfilled. According to OECD, IBM, and CSIRO of Australia, trust elements consist of explainability, fairness, accountability, robustness, transparency, and human-centered values. Products to achieve trust through enabling explainability, fairness, accountability, robustness, and transparency are widely researched, experimented with, and developed, especially under distributed system settings. Design patterns can be applied to achieve trust in software products by addressing recurring problems and increasing software quality attributes. The achieved trust can be extended to a broader software architecture if the product is utilized as a software component to collaborate with other components.



\subsection{Achieving Distributed Trust Through Process}

In recent years, multiple government bodies or NGOs have expressed their interest in regulating AI by providing multiple trustworthy, responsible, or ethical AI standards. The main motivations are to ensure that human-centered values and human interests are well preserved while utilizing AI on daily basis. For instance, the introduction of ChatGPT by OpenAI, GitHub Copilot, and many other AI services has garnered an extensive amount of attention on possible businesses ideas and values they may bring, to the extent that people start to question other than the benefits these AI services bring, are these AI services and the results they generated trustworthy and aligned with the human-centered values? In 2021, OECD produces a report entitled ``Tools for trustworthy AI - A framework to compare implementation tools for trustworthy AI systems~\cite{/content/paper/008232ec-en}'' that presents a framework for comparing tools and practices to implement trustworthy AI systems as set out in the OECD AI Principles. The framework aims to help collect, structure, and share information, knowledge, and lessons learned to date on tools, practices, and approaches for implementing trustworthy AI. The Australian government published the ``Australia’s AI Ethics Principles'' under Australia’s Artificial Intelligence Ethics Framework~\cite{doi_2022}. The framework outlined Australia’s eight AI ethics principles designed to ensure AI is safe, secure, and reliable. The framework aims to (1) achieve safer, more reliable, and fairer outcomes for all Australians, (2) reduce the risk of negative impact on those affected by AI applications, and (3) for businesses and governments to practice the highest ethical standards when designing, developing and implementing AI. 

Based on this framework and standards, risk assessments of AI systems on their trustworthiness, and ethics, have been widely conducted. For instance, Lu et al. propose a template for the AI system development process that enables AI ethics principles based on Australia's AI Ethics Framework \cite{lu2021software}. The template helps operationalize ethical AI in the form of concrete patterns and suggests a list of patterns using the newly created template. In 2022, the EU-US TTC (Trade and Technology Council) Joint Roadmap aims to advance shared terminologies and taxonomies, and inform their approaches to AI risk management and trustworthy AI on both sides of the Atlantic~\cite{ttc_2022}. Efficient and comprehensive governance solutions throughout the software development lifecycle can help improve the trustworthiness of blockchain systems for AI-based systems and data transaction systems. Similarly, Lu et al. proposed a Responsible AI (RAI) pattern catalogue{[\footnote{\url{https://research.csiro.au/ss/science/projects/responsible-ai-pattern-catalogue/}}]} that covers various architecture and governance patterns to enable trustworthiness principles and empower RAI systems.


In addition, researchers have approached blockchain governance from different angles. For example, Beck et al. \cite{selected14} highlighted the three major dimensions of IT governance, decision rights, incentives, and accountability, to blockchain governance. Allen and Berg \cite{selected7} explored exogenous and endogenous governance methods for blockchain platforms, while John and Pam \cite{selected10} and Pelt et al. \cite{selected11} investigated blockchain governance in on-chain and off-chain development processes. Tan et al. \cite{TAN2022101625} discussed blockchain governance decisions at three different levels, and Mosley et al. \cite{MOSLEY2022100085} and Dimitri \cite{info13060305} analyzed voting processes in blockchain systems. Liu et al. proposed a set of blockchain governance solutions as the structures and processes to elucidate who is involved and when to apply the solutions in blockchain systems \cite{pattern_language}. While in section~\ref{sec:blockchain_governance}, a comprehensive blockchain governance framework consisting of six high-level principles is introduced, to explain how governance is realised in blockchain systems regarding the decentralization level, decision rights, accountability, incentives, stakeholders, ecosystem, lifecycles, and legal and ethical responsibilities.

Ultimately, both product and process approach to achieving trust are closely related and complement one another as they are practically assessed and validated on their compatibility and fulfillment of the various non-functional requirements contributing to the trustworthiness of the system.

\section{Distributed Trust in AI-based Systems}

Trust is always about trusting in something. Being a distributed system that essentially handles data and trains AI models in distributed manners, and finally generates predictions or forecasts, the governance of trust for distributed AI systems highly relies on the fulfillment of trust requirements across multiple stages of AI model training and across the different stakeholders and nodes that interacts with each other. Furthermore, the products of the systems: (1) the AI model and (2) prediction results need to be trustworthy for the system to be deemed trustworthy. 

To effectively design software systems that fulfill intended trust qualities or requirements, software engineers and architects design software architectures by articulating components that address trust challenges. For instance, Lu et al. generated a collection of system-level design patterns to build trustworthy and responsible patterns into final AI products \cite{10007631}. The patterns are essentially mapped the responsible AI assurance patterns to the components of the reference architecture and discussed the trust and trustworthiness qualities fulfillment of each pattern \cite{lu2021software}. Moreover, federated learning has been deeply researched and applied thanks to its strength in data privacy preservation. Being a large-scale distributed system with different stakeholders, distributed trust challenges exist for federated learning to be widely adopted. Hence, various architectural research to enable distributed trust have been proposed. In this section, we will explore the different system-level solutions, including architecture designs and patterns to enable distributed trust for AI systems.

\subsection{Achieving Distributed Trust in AI-based Systems Through Product}

\subsubsection{Trustworthiness-by-design Architectural Patterns:}

\begin{figure}
\includegraphics[width=\textwidth]{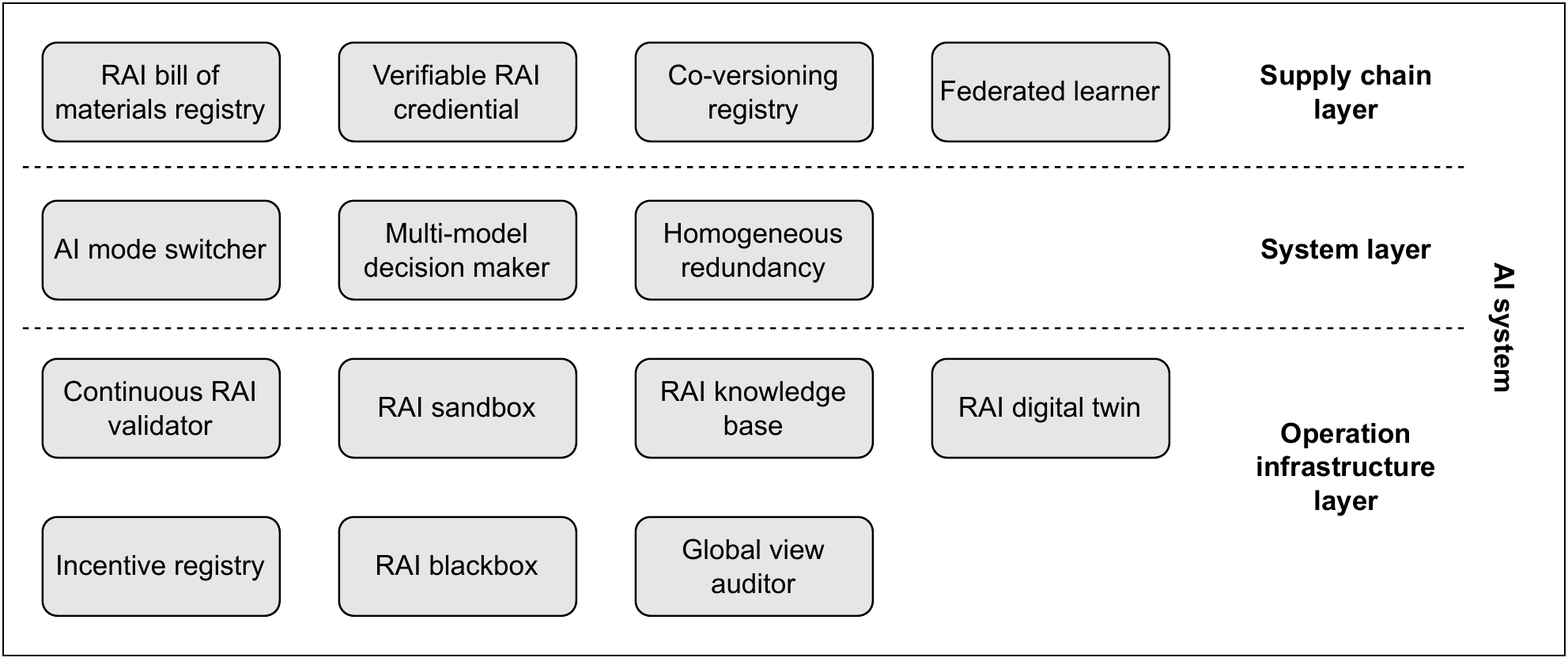}
\caption{Product patterns for responsible-AI-by-design per layers. \cite{raipatterns_lu23}.} \label{fig:AI_layers}
\end{figure}

\begin{figure}
\includegraphics[width=\textwidth]{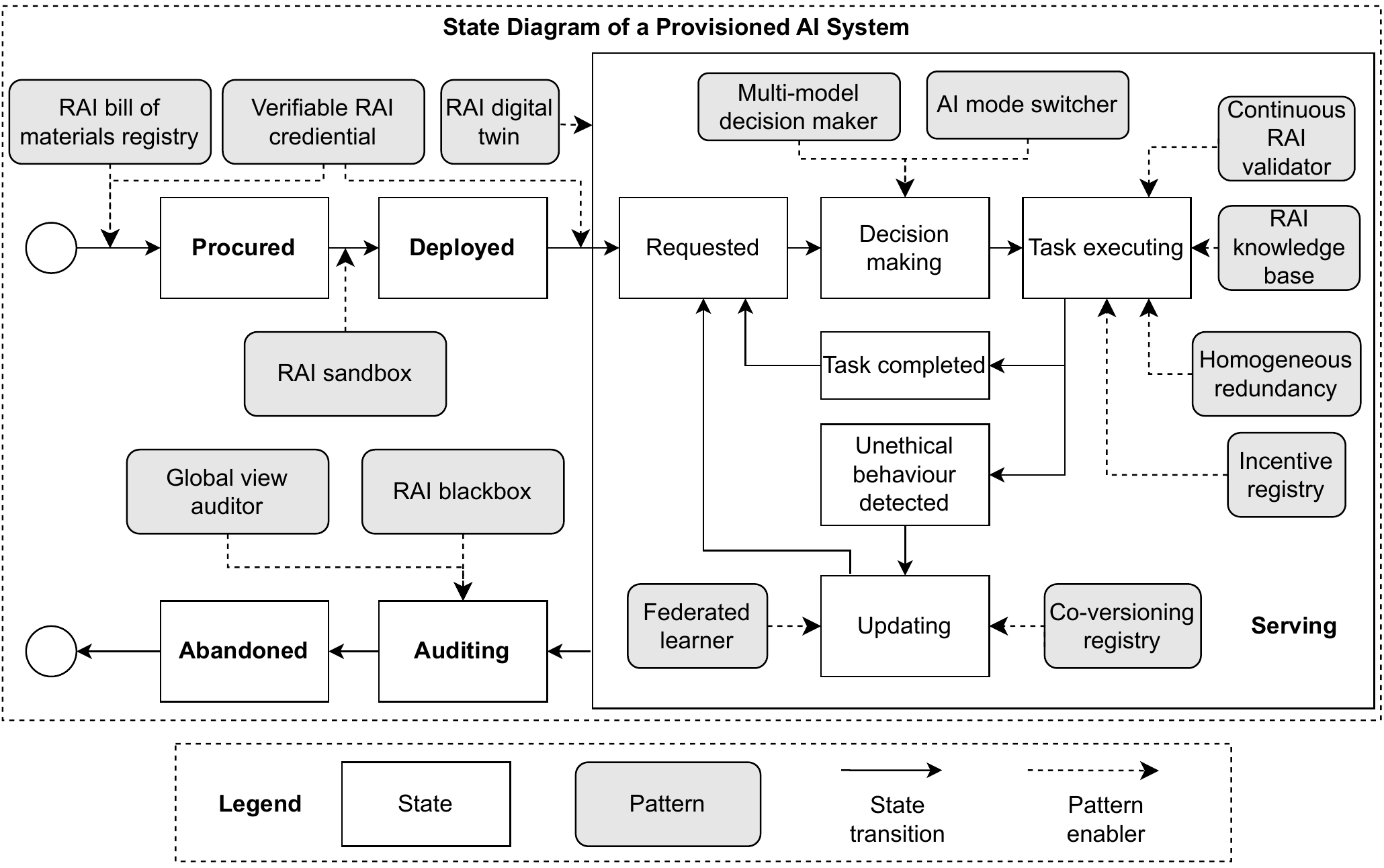}
\caption{Product patterns for responsible-AI-by-design \cite{raipatterns_lu23}.} \label{fig:state_diagram}
\end{figure}

\begin{figure}
\includegraphics[width=\textwidth]{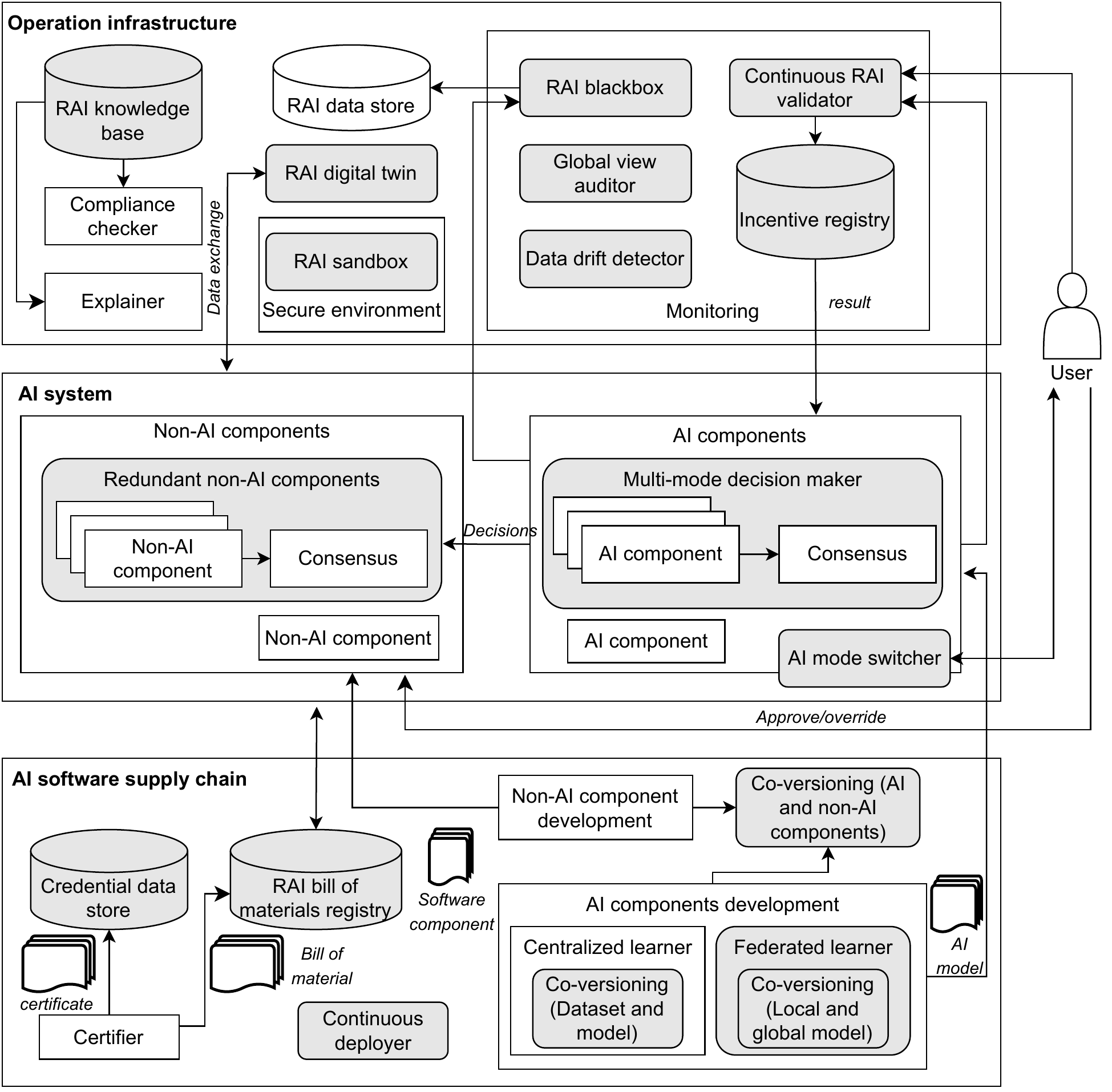}
\caption{Pattern-oriented responsible-AI-by-design reference architecture \cite{raipatterns_lu23}.} \label{fig:raira}
\end{figure}

Trustworthiness-by-design architectural patterns are set to provide system-level guidance on how to design the architecture of trustworthy and responsible AI systems. AI systems are comprised of three layers: (1) the supply chain layer that generates the software components which compose the AI system, (2) the system layer which is deployed the AI system, and (3) the operation infrastructure layer that provides auxiliary functions to the AI system. Fig. \ref{fig:AI_layers} presents the three layers and the product patterns under each of them. These product patterns can be embedded into AI systems as product features. Fig. \ref{fig:state_diagram} illustrates a state diagram of a provisioned AI system and highlights the patterns associated with relevant states while Fig. \ref{fig:raira} displays a pattern-oriented responsible-AI-by-design reference architecture.

The supply chain patterns include patterns that empower the trustworthiness of the supply chain components of AI systems. For instance, \textit{RAI bill of materials registry, verifiable RAI credential}, and \textit{co-versioning registry} enable the traceability of models and components of a system, whereas the \textit{federated learner} increases data privacy through federated learning. The system patterns deal with the systematic operation of the AI systems, including (1) \textit{AI mode switcher}, (2) \textit{multi-model decision maker}, and (3) \textit{homogeneous redundancy} that mainly increase human control. The operation infrastructure patterns include (1) \textit{continuous RAI validator}, (2) \textit{RAI sandbox}, (3) \textit{RAI knowledge base}, (4) \textit{RAI digital twin}, (5) \textit{incentive registry}, (6) \textit{RAI black box}, and (7) \textit{global view auditor} that mainly enhances the transparency, fairness, and traceability of the system from the infrastructure level.

\subsubsection{Trust Architecture and Patterns for Federated Learning Systems:}

\begin{figure}
\includegraphics[width=\textwidth]{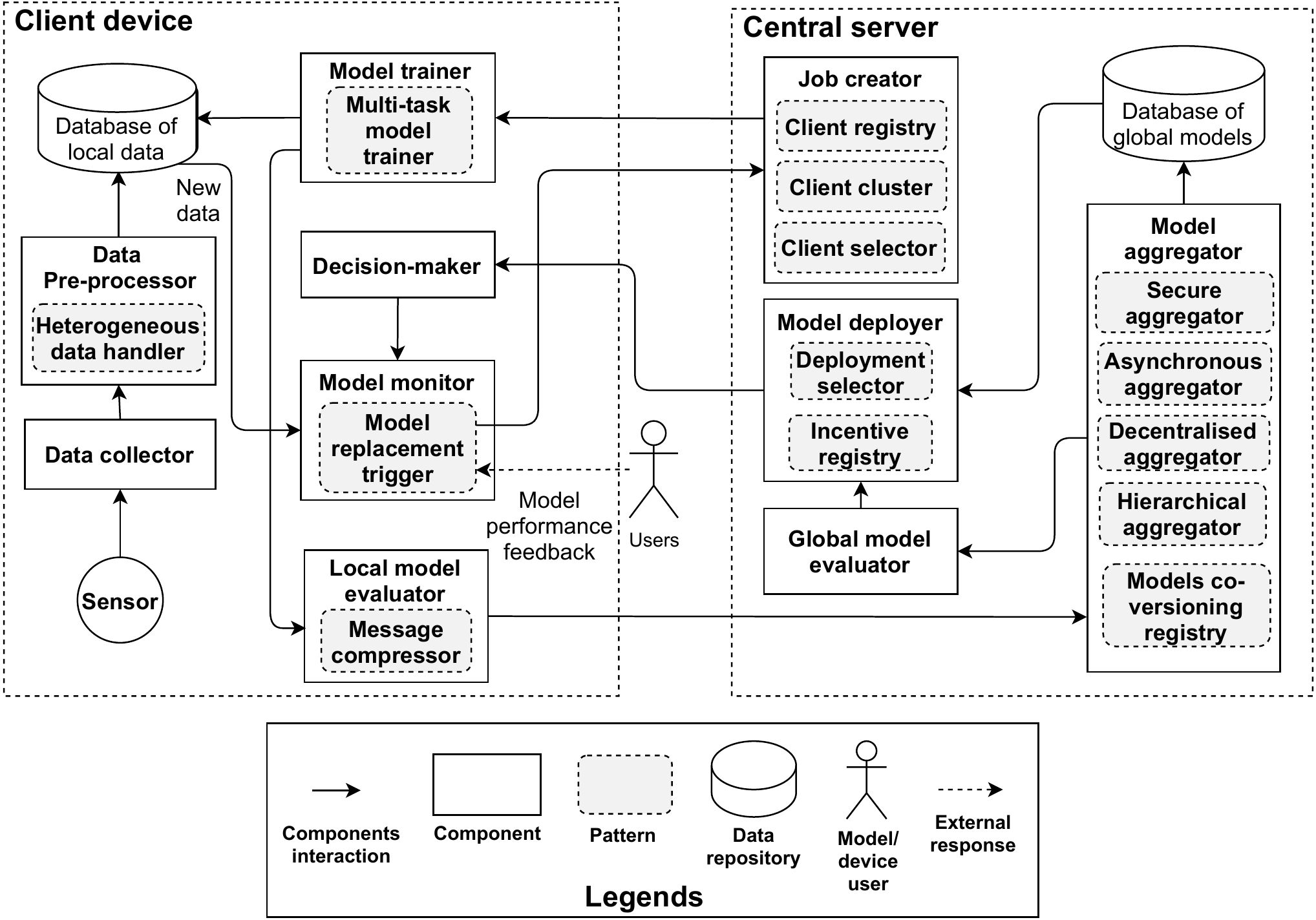}
\caption{FLRA reference architecture. \cite{10.1007/978-3-030-86044-8_6}} \label{FL_RA}
\end{figure}

Lo et al. proposed a pattern-oriented reference architecture generated based on a systematic literature review on federated learning, together with some of the industry best practices on machine learning system implementations \cite{10.1007/978-3-030-86044-8_6}. The reference architecture intends to provide an architecture design guideline and facilitate the end-to-end development and operations of federated learning systems while eliciting the different quality attributes and constraints to be considered. Fig.~\ref{FL_RA} illustrates the reference architecture accommodating the different components and patterns that are adaptable to enhance selective quality attributes of the system. 

There are several patterns that target to address the distributed trust challenges. For instance, the \textit{incentive registry} pattern increases the fairness of the model generated through incentive provisioning to the participating nodes. The \textit{model co-versioning registry} increases the traceability of the data and model version across multiple devices and the central server.

\subsubsection{Distributing Federated Learning via Blockchains:}

The adoption of blockchain as the digital infrastructure to realize application systems has been widely studied. The decentralized and immutability of blockchain increases the reliability and transparency of the system, which also empowers distributed trust. This section introduces several examples leveraging blockchain as a trust infrastructure in federated learning systems. 

\begin{figure}
\includegraphics[width=\textwidth]{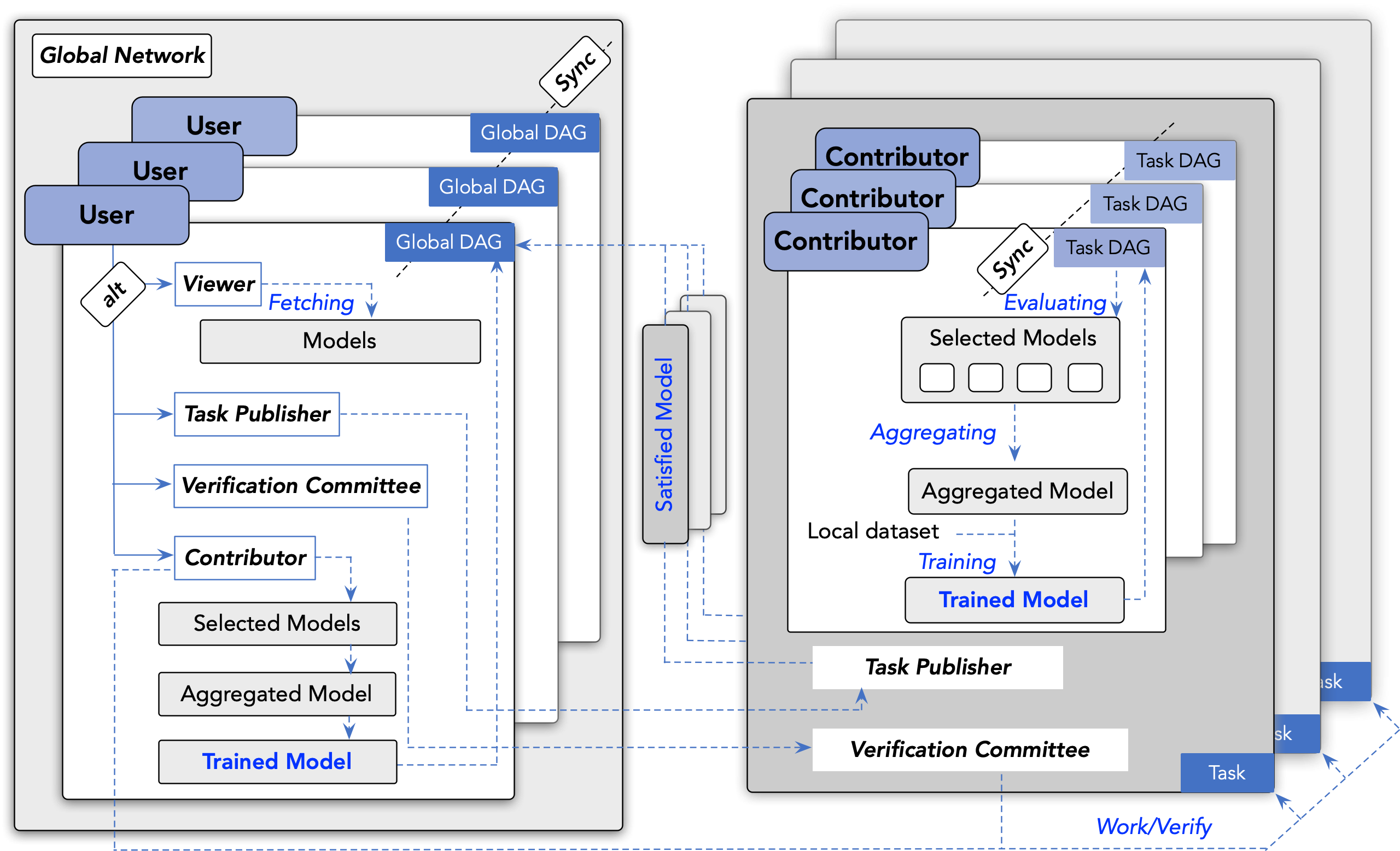}
\caption{The system archtecture of \textsc{IronForge}: a novel distributed FL framework \cite{ironfoge}.} 
\label{fig:ironforge}
\end{figure}

The IoTA project~\cite{iota} has set a precedent for boosting the throughput and scalability of blockchains by making the use of directed acyclic graph structure in blockchains, and has subsequently spawned many noteworthy projects such as Hashgraph~\cite{hashgraph}, Spectre~\cite{spectre}, Phantom~\cite{phantom}, Conflux~\cite{conflux}, etc. \textsc{IronForge}, inspired by the DAG-based incentive mechanism of IoTA, is the latest distributed FL system that features a DAG-based network structure to tackle the inconsistency in the decentralized FL process, excessive reliance on central coordination, and ineffective motivation of contributing to the learning resources at the same time. Specifically, \textsc{IronForge} builds a hybrid architecture (cf. Fig.~\ref{fig:ironforge}) that involves two types of 
DAG, namely, \textit{Task-DAG} in which users who aim at improving their local model prediction accuracy by virtue of the computational powers and resources of others can release tasks. and \textit{Global-DAG}, viewed as a ``unique'' and public model resource pool, in which users can hunt for models that uniquely meet their own local testing dataset. 
The training processes in both \textit{Task-DAG} and \textit{Global-DAG} are traceable owing to the DAG data structure. A DAG node published by a participant consists of a model update and the directed edges of the node indicate the aggregating relationship with existing models during the update, hence no central coordinator is required to conduct the training processes, and the distributed trust is achieved.

\begin{figure}
\includegraphics[width=\textwidth]{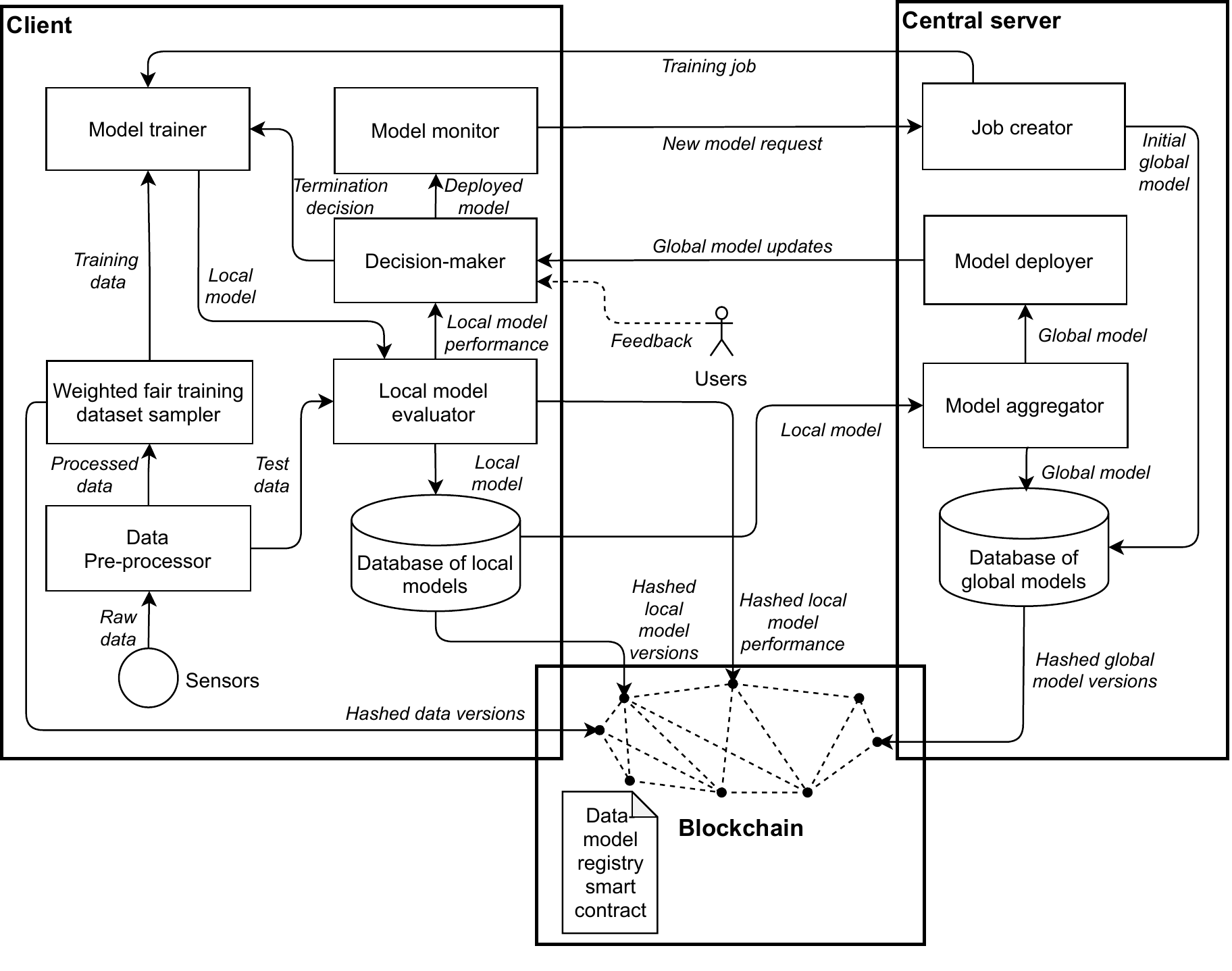}
\caption{Blockchain-based Trustworthy Federated Learning Architecture. \cite{9686048}} \label{FL_trust}
\end{figure}

Lo et al. proposed a trustworthy federated learning architecture design that focused on addressing two main aspects that enable trustworthiness in the federated learning architecture: (1) fairness, and (2) accountability. Fig. \ref{FL_trust} describes the blockchain-based trustworthy federated learning architecture. To enhance the fairness of the model prediction, a weighted fair training dataset sampler component is embedded to help minimize the bias and skewness in local training datasets. Then, the blockchain and smart contract-based data-model registry are adopted for data and model provenance, increasing the traceability and auditability of the data and model versions. Hence, the system can have better accountability.

\begin{figure}
\includegraphics[width=\textwidth]{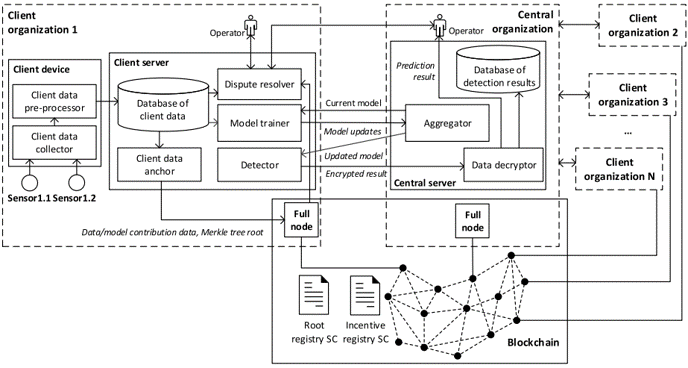}
\caption{Blockchain-based Federated Learning Architecture for Industrial IoT \cite{9233457}} \label{FL_iot}
\end{figure}

In additon, a blockchain-based federated learning system for failure detection in IIoT is proposed by Zhang et al., which enables verifiable integrity of client data \cite{9233457}. In the architecture, as shown in Fig. \ref{FL_iot}, each client periodically creates a Merkle tree in which each leaf node represents a client data record, and stores the tree root on a blockchain. This enables the traceability of data sources and since the records are immutable, accountability is ensured. Furthermore, a smart contact-based incentive mechanism is designed to improve the motivation of participation and fairness of the overall system in model training.

\subsection{Achieving Distributed Trust in AI-based Systems Through Process}

The process patterns to enable distributed trust of AI systems can be categorized into three main layers: (1) multi-level governance, (2) process-oriented, and (3) trustworthiness-by-design, each being responsible for establishing multi-level governance for responsible and trustworthy AI; setting up trustworthy development processes; or building trustworthy AI by design \cite{raipatterns_lu23}.

\subsubsection{Multi-level Governance Patterns:}

\begin{figure}
\includegraphics[width=\textwidth]{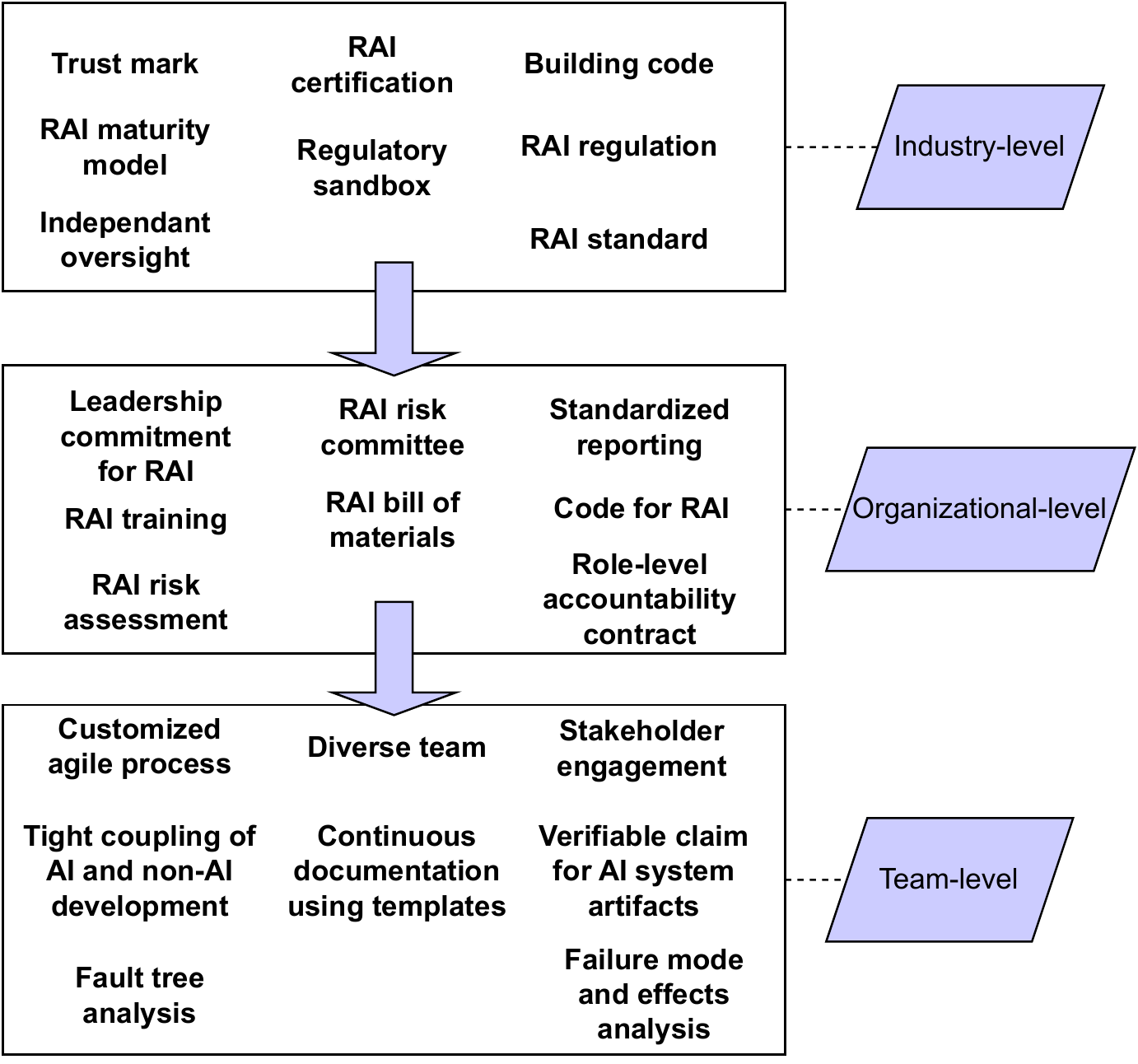}
\caption{Governance patterns per layer \cite{raipatterns_lu23}.} \label{fig:g_patterns}
\end{figure}

Firstly, the governance patterns are classified into industry-level governance patterns, organization-level governance patterns, and team-level governance patterns (refer Fig.~\ref{fig:g_patterns}). The target users of industry-level governance patterns are responsible AI (RAI) governors, while the impacted stakeholders include AI technology, solutions, and tools producers and procurers. For organization-level patterns, the target users are the management team and the impacted stakeholders are employees. The target users of team-level patterns are the development team \cite{raipatterns_lu23}. 

The industry-level governance patterns mainly assist in regulating and governing the AI applications or systems for an entire industry. Regulations and standards are set to oversee the trustworthiness of AI systems and services. For instance, \textit{AI regulation} assesses whether the software system would benefit from the incorporation of AI or whether it may be negatively impacted. Similarly, the \textit{RAI certification} serves as evidence to recognize organizations or individuals that have the ability to develop or use an AI system in a responsible manner, or developed processes or AI system or component that is compliant with standards or regulations. The \textit{RAI standard} describes repeatable processes for creating trustworthy and responsible AI systems that ultimately facilitate jurisdictions' interoperability. 

The organization-level governance patterns induce trustworthy and responsible AI from an organizational perspective where cooperation between the organization members is crucial, especially the supervision and support from the leadership, plus the active execution from the engineers to preserve and maintain trustworthiness. For example, \textit{Leadership commitment for RAI} actively promotes a culture of RAI within the organization, provides training and guidelines on RAI practices, and establishes clear founding ethics principles and governance structure for the organization to nurture the importance of maintaining trustworthy and responsible AI systems and products. \textit{RAI risk committee} governs and monitors the establishment of standard processes for decision-making of AI projects within an organization. The \textit{RAI bill of materials} is a list of components used to create an AI software product.

Lastly, the team-level governance patterns deal directly with the development of AI systems or products, spanning across the development process, stakeholders' and developers' engagement, and verifiability of the artifacts. For instance, the \textit{customized agile process} explored agile methods to consider ethics and trust principles by adding extension points (artifacts, roles, ceremonies, practices, and culture) to the agile process of AI systems development. In addition, \textit{tight coupling of AI and non-AI development} pattern manages trust through close coordination of project deliverables and progress, including both AI components that produce AI models and non-AI components that utilize the outputs of the AI models for overall system functionalities. The \textit{verifiable claim for AI system artifacts} pattern enables verification mechanisms that allow users or developers to flag issues or provide experience reports and stakeholders could directly investigate an AI system’s ethical properties.

\subsubsection{Process-oriented Patterns for Trustworthiness:}

\begin{figure}
\includegraphics[width=\textwidth]{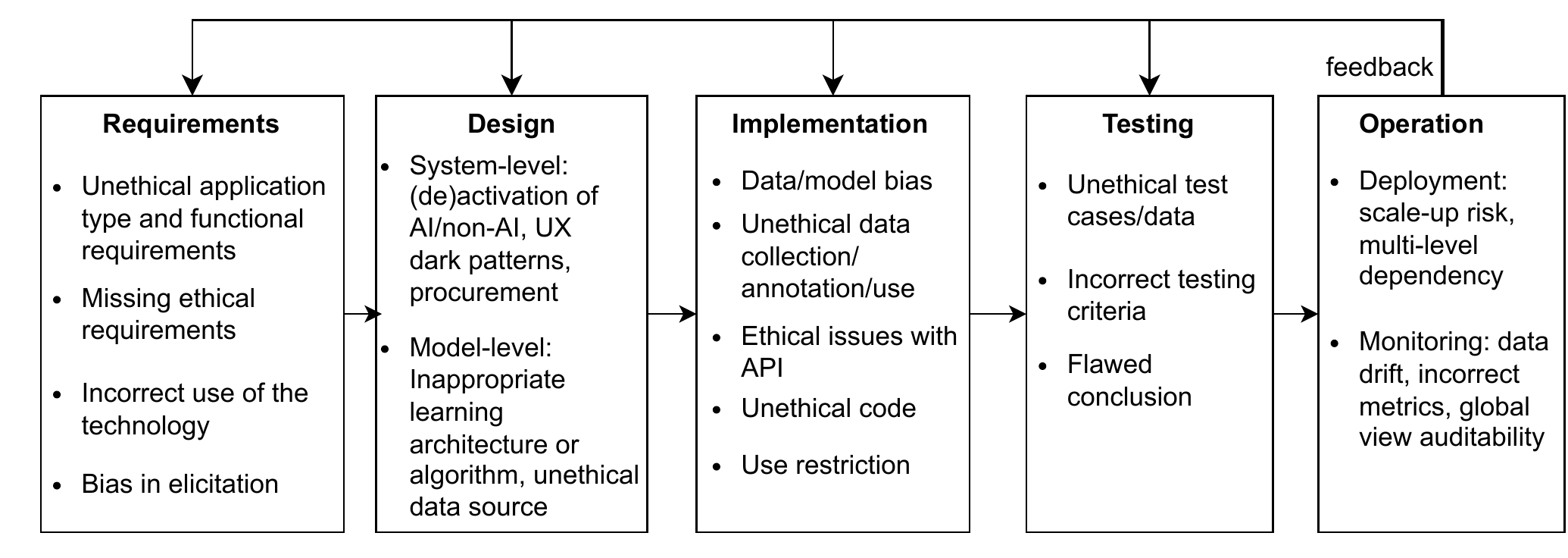}
\caption{AI system development lifecycle with potential trust risks \cite{raipatterns_lu23}.} \label{fig:risk_patterns}
\end{figure}

\begin{figure}
\includegraphics[width=\textwidth]{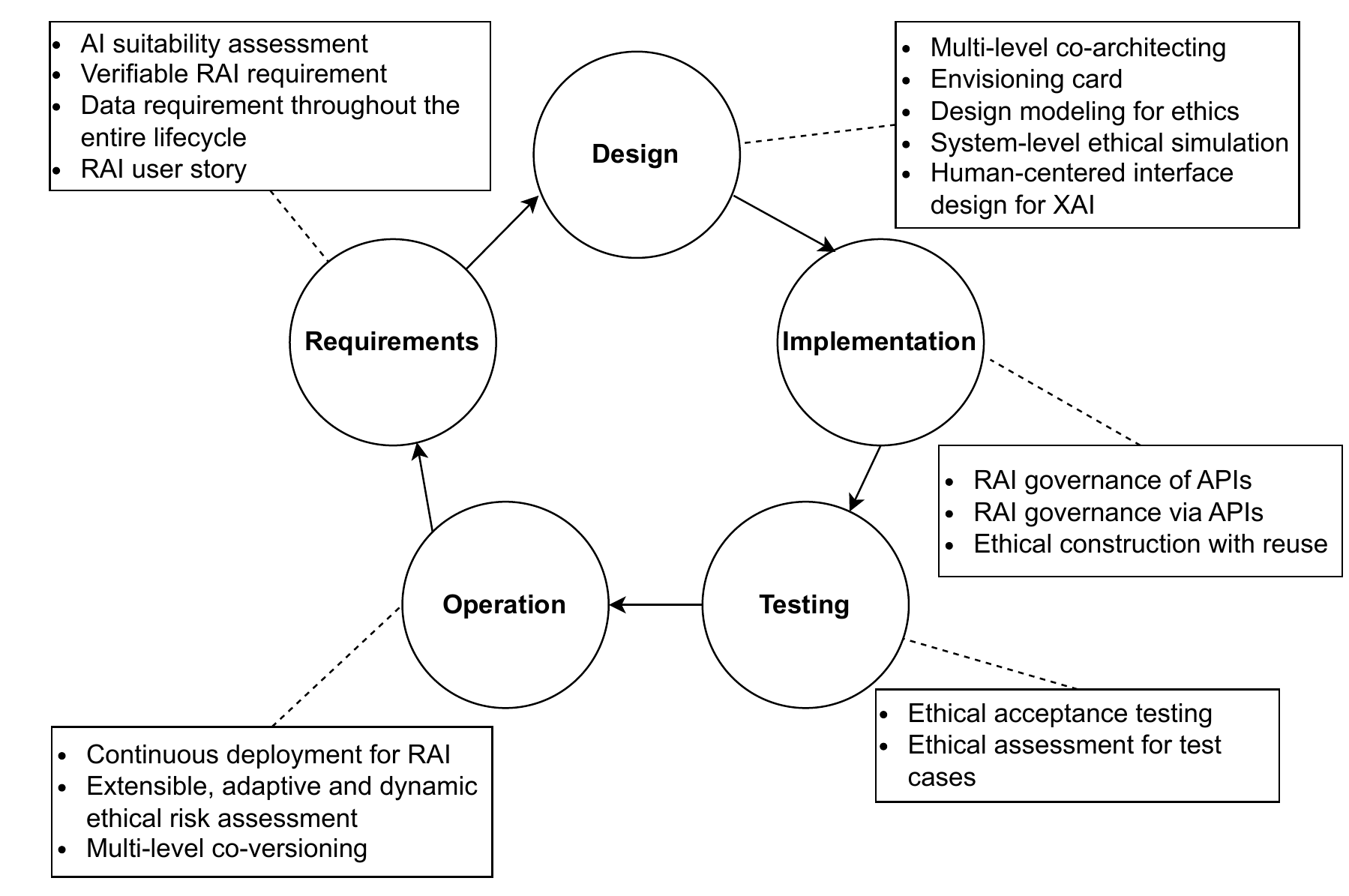}
\caption{Process patterns under development lifecycle \cite{raipatterns_lu23}.} \label{fig:process_patterns}
\end{figure}

The process patterns mainly cover the patterns that are applicable in each AI system development stage. Fig.~\ref{fig:risk_patterns} illustrates the stages of the system development process lifecycle and the potential ethical and trust risks under each stage, and Fig.~\ref{fig:process_patterns} shows the process patterns applicable under each stage.

The patterns for the requirement stage are: (1) \textit{AI suitability assessment}, (2) \textit{verifiable RAI requirement}, (3) \textit{lifecycle-driven data requirement}, and (4) \textit{RAI user story}. The \textit{suitability assessment} assesses whether the software system would benefit from the incorporation of AI and that it has high suitability and transparency. \textit{verifiable RAI requirement} pattern documents the RAI requirements in a clear and verifiable with specific acceptance criteria to reduce ethical or trust risk. The \textit{lifecycle-driven data requirement} ensures that data is used and managed in a responsible manner, explicitly listing and specifying data requirements throughout the entire data lifecycle, including collection, cleaning, preparation, validation, analysis, and termination. These requirements should take into account all relevant ethical principles and all stakeholders to ensure traceability and accountability. The \textit{RAI user story} is integrated into the agile development process to ensure traceability and elicitation of RAI requirements.

The patterns for the design stage are: (1) \textit{multi-level co-architecting}, (2) \textit{envisioning card}, (3) \textit{RAI design modelling}, (4) \textit{system-level RAI simulation}, and (5) \textit{XAI interface}. The \textit{multi-level co-architecting} pattern handles the seamless integration of different components of multi-level co-architecting, including co-designing both AI and non-AI components, as well as co-designing different components within the AI model pipeline. The \textit{envisioning card} pattern is created to assist the development team in incorporating human values into the design processes of AI systems. \textit{RAI design modeling} pattern models the architecture of AI systems and represents their ethical aspects, designing formal models that take into account human values, using ontology to model AI system artifacts for accountability, creating RAI knowledge bases to inform design decisions that consider ethical concerns, and using logic programming to implement ethical principles. The \textit{system-level RAI simulation} pattern comprehends the characteristics and behaviors of AI systems through simulation to assess potential RAI risks before deploying them in the real world. \textit{XAI interface} focuses on human-AI interaction problems to achieve human-centered interface design.

The patterns for the implementation stage are: (1) \textit{RAI governance of APIs}, (2) \textit{RAI governance via APIs}, and (3) \textit{RAI construction with reuse}. The \textit{RAI governance of APIs} and \textit{RAI governance via APIs} check the compliance of APIs of AI systems and provide AI services on the cloud and control the interactions with these services via APIs to achieve trust. The \textit{RAI construction with reuse} pattern facilitates the trading of reusable AI assets, including component code, models, and datasets. Blockchain technology can be utilized to create an immutable and transparent marketplace, allowing the auction-based trading of AI assets and material assets, such as cloud resources. The patterns for the testing stage: (1) \textit{RAI acceptance testing}, and (2) \textit{RAI assessment for test cases}. Both patterns actively conduct tests to determine if the ethical requirements of an AI system are met and if their development processes are trustworthy and responsible. 

Lastly, the patterns for the operation stage are: (1) \textit{continuous deployment for RAI}, (2) \textit{extensible, adaptive and dynamic RAI risk assessment}, and (3) \textit{multi-level co-versioning}. The \textit{continuous deployment for RAI} seamlessly deploys the AI models into production environments by utilizing various deployment strategies while ensuring the fulfillment of RAI requirements. The \textit{extensible, adaptive and dynamic RAI risk assessment} pattern continuously performs risk assessment and mitigation for RAI systems. The \textit{multi-level co-versioning} supports the model versioning of multi-level or multi-party systems (for e.g., federated learning). 

\subsection{Achieving Distributed Trust in Data and Transaction}


Blockchain technology is considered to provide a digital infrastructure for trustless interactions between stakeholders. Untrusted nodes can reach consensus on transactional data states via predefined algorithmic rules, hence trusted third-parties are removed from on-chain business processes \cite{scheuermann2015iacr}. Nevertheless, the inherent immutability characteristic and code-based algorithmic governance rules have increased concerns about whether blockchain systems are designed, developed and operated in a trustworthy way, especially after the hard forks of the most two well-known blockchain systems, Bitcoin and Ethereum. 

There was a tedious dispute about Bitcoin block size endured from 15 August 2015 to 23 January 2016, resulting in the split of Bitcoin community and ecosystem \cite{selected20}. Ethereum's Decentralized Autonomous Organisation (DAO) was hacked due to the vulnerabilities in smart contract source code, which caused the loss of over 60 million US dollars. The remedy was to rollback the Ethereum blockchain to a specific block height before the attack, but a small part of the community decide to continue the original Ethereum blockchain, which ended up as two separate blockchain systems \cite{DAOattack}. 

These two crises raised awareness and interest in enhancing distributed trust in blockchain systems. Specifically, this can be achieved via both product and process approaches. In this section, we discuss achieving distributed trust in blockchain and blockchain-based applications from the perspective of software architecture, including blockchain design patterns, leveraging blockchain as a software connector, blockchain governance, Decentralized Autonomous Organizations, and blockchain-based federated learning.

\subsection{Achieving Distributed Trust in Blockchain Through Product: Patterns and Architecture}
\label{sec:blockchain_pattern}

A pattern collection\footnote{\url{https://research.csiro.au/blockchainpatterns/}} was proposed to assist system architects and developers in addressing the recurring problems during the design of a blockchain and its application systems, by providing systematic and holistic guidance on the applicable reusable solutions. Each pattern is to resolve a certain problem, which can improve particular software quality attributes and improve the trustworthiness of blockchain. The pattern collection consists of general design patterns for blockchain-based applications \cite{bc_pattern}, data migration \cite{migration_pattern}, and application-specific design patterns for payment \cite{payment_pattern} and self-sovereign identity (SSI) applications \cite{ssi_pattern}. We summarize the extant patterns as follows.

\begin{itemize}
    \item \textit{Data management patterns}: Considering the requirements of performance, integrity, and privacy, and the application scenario of data volume, velocity, and variety when determining whether to store certain data on-chain.

    \item \textit{Interact with external world patterns}: Communicating with other off-chain components in a large software system.

    \item \textit{Security patterns}: Ensuring the security of smart contract interactions.

    \item \textit{Smart contract patterns}: Reducing the cost of designing, deploying, and using smart contracts in terms of the time, space, and message complexity.

    \item \textit{Deployment patterns}: Simplifying the interactions between users and blockchain-based applications.

    \item \textit{Data migration patterns}: Facilitating the business process, performance, cost efficiency, privacy, and regulatory compliance by migrating data from one blockchian system to another.

    \item \textit{Payment patterns}: Enabling the state transitions of on-chain tokens for blockchain-based payment applications.

    \item \textit{SSI patterns}: Managing the on-chain keys, decentralized identifiers, and credentials for better usage of SSI applications.
\end{itemize}

\begin{figure}
\centering
\includegraphics[width=0.5\textwidth]{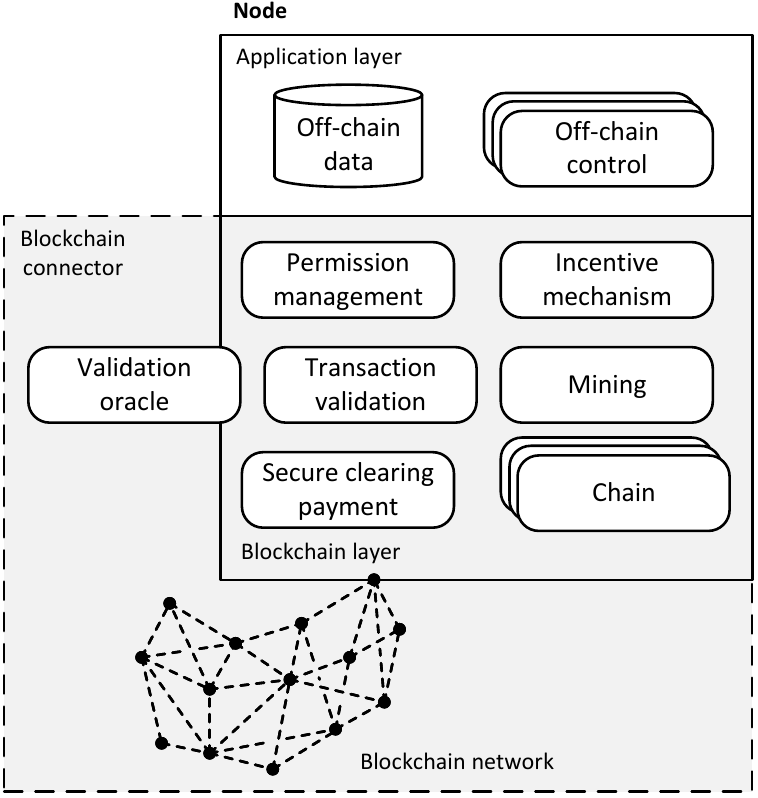}
\caption{Overview of blockchain as connector \cite{connector}.} \label{fig:connector}
\end{figure}

As illustrated in Fig.~\ref{fig:connector}, blockchain can be utilized as a software connector to provide services for communication, coordination, and facilitation \cite{connector}. First, other software components can transfer data using blockchain as a meidator, via both sending messages through transactions, and storing certain data in smart contract storage. Secondly, blockchain smart contracts provide coordination services for the computation of other components. Thirdly, the facilitation services include the choice of transaction validation and block mining approaches (e.g., calculating random data in ``Proof-of-Work'', possessing tokens in ``Proof-of-Stake'', etc.), secure payment through cryptography, economic incentives, reputation and rating mechanisms for reward and assess the performance of miners, to name a few. Applying design patterns can achieve distributed trust in blockchain, as software quality attributes (e.g., security, scalability) are improved. Further, leveraging blockchain as a connector in software architecture design can help achieve distributed trust in application systems, since blockchain provides an underlying distributed infrastructure where trusted third-parties can be eliminated.

\begin{figure}
\includegraphics[width=\textwidth]{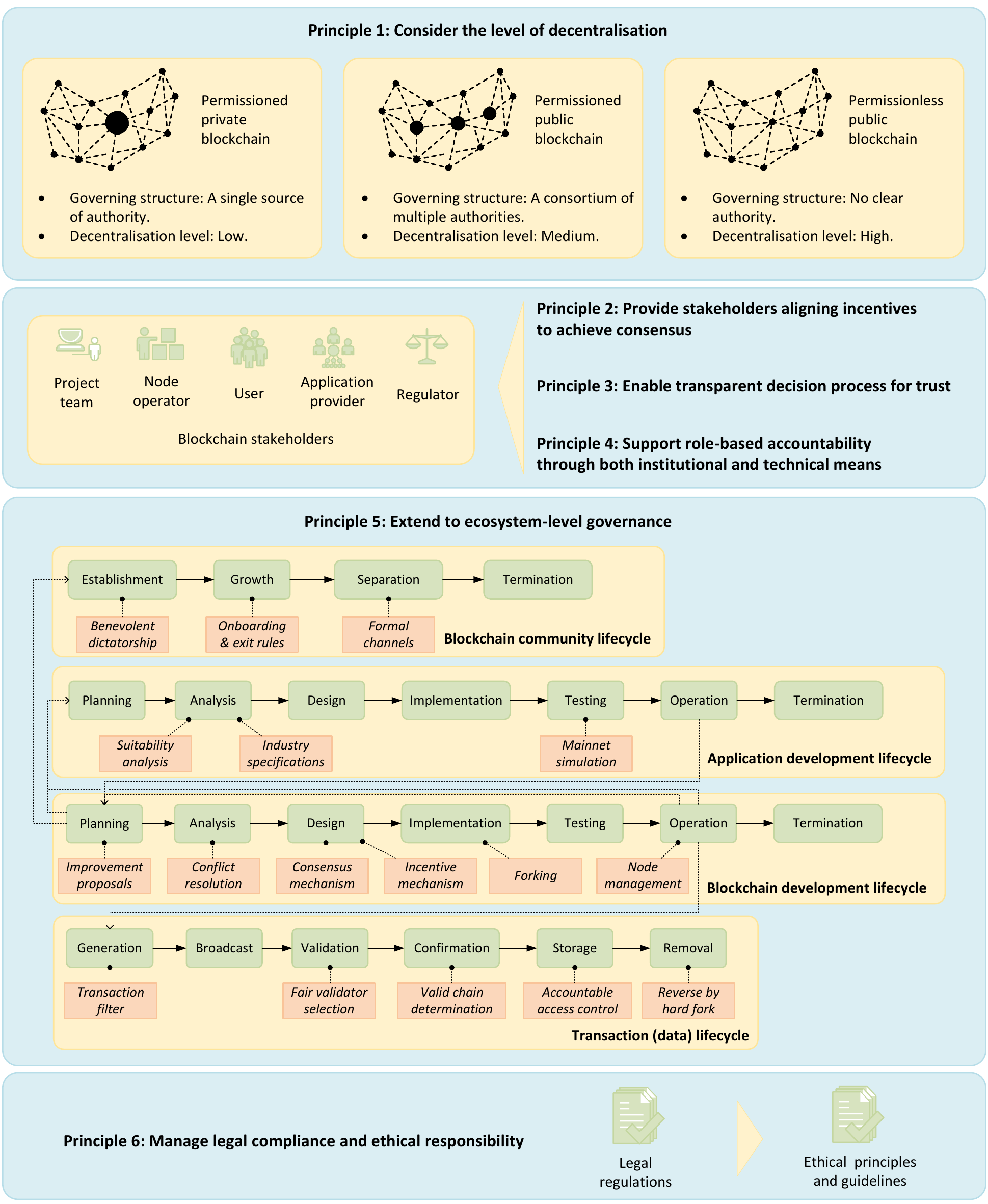}
\caption{Blockchain governance framework \cite{liu2021defining}.} \label{fig:framework}
\end{figure}

\subsection{Achieving Distributed Trust in Blockchain Through Product: Governance}
\label{sec:blockchain_governance}

Blockchain governance refers to the structures and processes that are designed to ensure the development, use, and maintenance of blockchain are compliant with legal regulations and ethical responsibilities \cite{liu2021systematic}. Based on the results of a systematic literature review, Liu et al. proposed a blockchain governance framework consisting of six principles \cite{liu2021defining}, as demonstrated in Fig.~\ref{fig:framework}. 

First, blockchain can be categorized into three different types for diverse requirements. The decentralization level of each blockchain type can impact the governance structure in terms of the allocation of decision rights, accountability, and incentives. First, in order to gain the trust of all stakeholders, it is crucial for blockchain systems to have a transparent decision-making process, which helps oversee whether governance decisions are reasonable. Secondly, the establishment of accountability can be accomplished via institutional and technical means to ensure stakeholders can be identified and held answerable for their decisions. Thirdly, Within blockchain governance, incentives play a significant role in driving stakeholder behaviour. Providing incentives can both motivate desirable behaviours and resolve conflicts in a blockchain system.

In addition, the implementation of blockchain governance should be extended across the entire ecosystem. For on-chain transactions, governance emphasizes the importance of accountable access control. Depending on the deployed blockchain type, the capabilities of sending, validating, and reading transactions are assigned to different stakeholders. In the case of blockchain platforms, formalized procedures must be followed to finalize improvement proposals. Furthermore, any changes to blockchain-based applications must comply with industry regulations and specifications, which may require upgrades to the underlying blockchain platform. In off-chain community governance, stakeholders are divided into various groups based on their roles and decision-making authority, and they may have different communication channels for certain issues.

Note that blockchain governance needs to ensure that related decisions and processes conform to legal regulations and ethical responsibilities. To manage legal compliance, local and international policies must be adhered to. To promote ethical guidelines, human values must be preserved within blockchain governance.

\subsection{Exploiting Automated Organization via Distributed Trust: DAO}

DAOs are a form of blockchain-based organization that operates without the need for a central authority or intermediary. Decisions are made through a decentralized group of users that collectively hold decision-making power, with transactions and operations recorded on the underlying ledgers. This makes DAOs a prime example of the effectiveness of distributed trust, with the system's trust being upheld by the consensus algorithms and transparency of the blockchain network.

The core of a DAO has been well-defined~\cite{dao,dao2}, and is mainly made up of the following components.
\begin{itemize}
    \item A set of smart contracts that are hardcoded to enforce the rules and regulations of the DAO, and are executed automatically to govern the operation and decision-making processes under the nature of decentralization.

    \item A set of on-chain decentralized identifiers (DIDs) that enable users to create and manage their own identifiers, as well as to verify and authenticate others' identities in a secure and decentralized way.

    \item A set of wallet services in which stakes and utility tokens can be securely stored in order to offer DAO's services such as accessing to specific products and sub-services, establishing on-chain reputations, accumulating confidence and power for governance tasks (e.g., e-voting), etc. 
\end{itemize}
As such, distributed trust in DAOs is based on the security and immutability of blockchains, as well as the transparency and accountability of the governance mechanisms that determine its actions. By leveraging distributed trust, DAOs have the potential to alter the way organizations are formed, operated, and governed, which enables greater involvement, transparency, and innovation.

\section{Conclusions}

Trust architecture is becoming distributed and underpins everything. Especially with the emergence of various distributed and decentralized systems (i.e. federated learning, blockchain). In summary, we have described and defined the concept of distributed trust from multiple perspectives, clearly underlining the different trustworthiness principles. We have also identified the different system-level solutions and governance processes to enable distributed trust and visualized them under a multi-layer distributed trust architecture, covering the different components of a software system. We expressed the architecture-style solutions into products and the governance solution as the process of achieving distributed trust. Then, we classify the state-of-the-art solutions to achieve distributed trust for AI-based systems, and data and transaction systems. We have also described multiple applications of federated learning supported by blockchain to enable distributed trust. Simultaneously, we encapsulated the various governance processes that complement the design patterns and tactics, ranging from Responsible AI multi-level governance patterns to blockchain governance approaches. For future work, we intend to extrapolate the review to more algorithmic approaches to enable distributed trust.
%
%
%
\bibliographystyle{splncs04}
\bibliography{mybibliography}

\begin{thebibliography}{10}
\providecommand{\url}[1]{\texttt{#1}}
\providecommand{\urlprefix}{URL }
\providecommand{\doi}[1]{https://doi.org/#1}

\bibitem{10.1145/502585.502638}
Aberer, K., Despotovic, Z.: Managing trust in a peer-2-peer information system.
  In: Proceedings of the Tenth International Conference on Information and
  Knowledge Management. p. 310–317. CIKM '01, Association for Computing
  Machinery, New York, NY, USA (2001). \doi{10.1145/502585.502638},
  \url{https://doi.org/10.1145/502585.502638}

\bibitem{Aikebaier2011}
Aikebaier, A., Enokido, T., Takizawa, M.: Trustworthy group making algorithm in
  distributed systems  \textbf{1}(1), ~6. \doi{10.1186/2192-1962-1-6},
  \url{https://doi.org/10.1186/2192-1962-1-6}

\bibitem{selected7}
Allen, D.W., Berg, C.: {Blockchain Governance: What we can Learn from the
  Economics of Corporate Governance}. The Journal of The British Blockchain
  Association p. 12455 (2020)

\bibitem{mckinsey22}
analysis, M.: Trust architectures and digital identity

\bibitem{DAOattack}
Atzei, N., Bartoletti, M., Cimoli, T.: A survey of attacks on ethereum smart
  contracts (sok). In: Maffei, M., Ryan, M. (eds.) Principles of Security and
  Trust. pp. 164--186. Springer Berlin Heidelberg, Berlin, Heidelberg (2017)

\bibitem{hashgraph}
Baird, L.: The swirlds hashgraph consensus algorithm: Fair, fast, byzantine
  fault tolerance. Swirlds Tech Reports SWIRLDS-TR-2016-01, Tech. Rep
  \textbf{34},  9--11 (2016)

\bibitem{migration_pattern}
Bandara, H.D., Xu, X., Weber, I.: Patterns for blockchain data migration. In:
  Proceedings of the European Conference on Pattern Languages of Programs 2020.
  EuroPLoP '20, Association for Computing Machinery, New York, NY, USA (2020).
  \doi{10.1145/3424771.3424796}, \url{https://doi.org/10.1145/3424771.3424796}

\bibitem{bass2003software}
Bass, L., Clements, P., Kazman, R.: Software architecture in practice.
  Addison-Wesley Professional (2003)

\bibitem{selected14}
Beck, R., M{\"u}ller-Bloch, C., King, J.L.: Governance in the blockchain
  economy: A framework and research agenda. Journal of the Association for
  Information Systems  \textbf{19}(10), ~1 (2018)

\bibitem{10.1007/978-3-540-24769-2_14}
Bosch, J.: Software architecture: The next step. In: Oquendo, F., Warboys,
  B.C., Morrison, R. (eds.) Software Architecture. pp. 194--199. Springer
  Berlin Heidelberg, Berlin, Heidelberg (2004)

\bibitem{5072249}
Choi, H.L., Brunet, L., How, J.P.: Consensus-based decentralized auctions for
  robust task allocation. IEEE Transactions on Robotics  \textbf{25}(4),
  912--926 (2009). \doi{10.1109/TRO.2009.2022423}

\bibitem{ttc_2022}
Commission, E.: Ttc joint roadmap for trustworthy ai and risk management (Dec
  2022),
  \url{https://digital-strategy.ec.europa.eu/en/library/ttc-joint-roadmap-trustworthy-ai-and-risk-management}

\bibitem{selected20}
De~Filippi, P., Loveluck, B.: The invisible politics of bitcoin: governance
  crisis of a decentralized infrastructure. Internet Policy Review
  \textbf{5}(4) (2016). \doi{https://doi.org/10.14763/2016.3.427}

\bibitem{dean2008mapreduce}
Dean, J., Ghemawat, S.: Mapreduce: simplified data processing on large
  clusters. Communications of the ACM  \textbf{51}(1),  107--113 (2008)

\bibitem{info13060305}
Dimitri, N.: Quadratic voting in blockchain governance. Information
  \textbf{13}(6) (2022). \doi{10.3390/info13060305},
  \url{https://www.mdpi.com/2078-2489/13/6/305}

\bibitem{doi_2022}
Department~of Industry, S., Resources: Australia's ai ethics principles (Oct
  2022),
  \url{https://www.industry.gov.au/publications/australias-artificial-intelligence-ethics-framework/australias-ai-ethics-principles}

\bibitem{selected10}
John, T., Pam, M.: Complex adaptive blockchain governance. In: MATEC Web of
  Conferences. vol.~223, p. 01010. EDP Sciences (2018)

\bibitem{conflux}
Li, C., Li, P., Zhou, D., Yang, Z., Wu, M., Yang, G., Xu, W., Long, F., Yao,
  A.C.C.: A decentralized blockchain with high throughput and fast
  confirmation. In: 2020 $\{$USENIX$\}$ Annual Technical Conference
  ($\{$USENIX$\}$$\{$ATC$\}$ 20). pp. 515--528 (2020)

\bibitem{LI2018133}
Li, Z., Barenji, A.V., Huang, G.Q.: Toward a blockchain cloud manufacturing
  system as a peer to peer distributed network platform. Robotics and
  Computer-Integrated Manufacturing  \textbf{54},  133--144 (2018).
  \doi{https://doi.org/10.1016/j.rcim.2018.05.011},
  \url{https://www.sciencedirect.com/science/article/pii/S073658451830022X}

\bibitem{ssi_pattern}
Liu, Y., Lu, Q., Paik, H.Y., Xu, X.: Design patterns for blockchain-based
  self-sovereign identity. In: Proceedings of the European Conference on
  Pattern Languages of Programs 2020. EuroPLoP '20, Association for Computing
  Machinery, New York, NY, USA (2020). \doi{10.1145/3424771.3424802},
  \url{https://doi.org/10.1145/3424771.3424802}

\bibitem{pattern_language}
Liu, Y., Lu, Q., Yu, G., Paik, H.Y., Perera, H., Zhu, L.: A pattern language
  for blockchain governance. In: Proceedings of the 27th European Conference on
  Pattern Languages of Programs. EuroPLop '22, Association for Computing
  Machinery, New York, NY, USA (2023). \doi{10.1145/3551902.3564802},
  \url{https://doi.org/10.1145/3551902.3564802}

\bibitem{liu2021defining}
Liu, Y., Lu, Q., Yu, G., Paik, H.Y., Zhu, L.: Defining blockchain governance
  principles: A comprehensive framework. Information Systems p. 102090 (2022).
  \doi{https://doi.org/10.1016/j.is.2022.102090},
  \url{https://www.sciencedirect.com/science/article/pii/S0306437922000758}

\bibitem{liu2021systematic}
Liu, Y., Lu, Q., Zhu, L., Paik, H.Y., Staples, M.: A systematic literature
  review on blockchain governance. Journal of Systems and Software p. 111576
  (2022). \doi{https://doi.org/10.1016/j.jss.2022.111576},
  \url{https://www.sciencedirect.com/science/article/pii/S0164121222002527}

\bibitem{9686048}
Lo, S.K., Liu, Y., Lu, Q., Wang, C., Xu, X., Paik, H.Y., Zhu, L.: Toward
  trustworthy ai: Blockchain-based architecture design for accountability and
  fairness of federated learning systems. IEEE Internet of Things Journal
  \textbf{10}(4),  3276--3284 (2023). \doi{10.1109/JIOT.2022.3144450}

\bibitem{10.1007/978-3-030-86044-8_6}
Lo, S.K., Lu, Q., Paik, H.Y., Zhu, L.: {FLRA}: A reference architecture
  for federated learning systems. In: Software Architecture. pp. 83--98.
  Springer International Publishing (2021)

\bibitem{lo2021architectural}
Lo, S.K., Lu, Q., Zhu, L., Paik, H.Y., Xu, X., Wang, C.: Architectural patterns
  for the design of federated learning systems. Journal of Systems and Software
   \textbf{191},  111357 (2022)

\bibitem{payment_pattern}
Lu, Q., Xu, X., Bandara, H.D., Chen, S., Zhu, L.: Patterns for blockchain-based
  payment applications. In: 26th European Conference on Pattern Languages of
  Programs. EuroPLoP'21, Association for Computing Machinery, New York, NY, USA
  (2022). \doi{10.1145/3489449.3490006},
  \url{https://doi.org/10.1145/3489449.3490006}

\bibitem{10007631}
Lu, Q., Zhu, L., Xu, X., Whittle, J.: Responsible-ai-by-design: A pattern
  collection for designing responsible ai systems. IEEE Software pp.~1--7
  (2023). \doi{10.1109/MS.2022.3233582}

\bibitem{raipatterns_lu23}
Lu, Q., Zhu, L., Xu, X., Whittle, J.: Responsible ai pattern catalogue (2023),
  \url{https://research.csiro.au/ss/science/projects/responsible-ai-pattern-catalogue/}

\bibitem{lu2021software}
Lu, Q., Zhu, L., Xu, X., Whittle, J., Douglas, D., Sanderson, C.: Software
  engineering for responsible ai: An empirical study and operationalised
  patterns. arXiv preprint arXiv:2111.09478  (2021)

\bibitem{mccurdy_2021}
McCurdy, C.:  (May 2021),
  \url{https://community.ibm.com/community/user/security/blogs/chris-mccurdy1/2021/05/19/ibm-security-and-microsoft-partnering}

\bibitem{MOSLEY2022100085}
Mosley, L., Pham, H., Guo, X., Bansal, Y., Hare, E., Antony, N.: Towards a
  systematic understanding of blockchain governance in proposal voting: A dash
  case study. Blockchain: Research and Applications  \textbf{3}(3),  100085
  (2022). \doi{https://doi.org/10.1016/j.bcra.2022.100085},
  \url{https://www.sciencedirect.com/science/article/pii/S2096720922000264}

\bibitem{/content/paper/008232ec-en}
OECD: Tools for trustworthy ai (312) (2021).
  \doi{https://doi.org/https://doi.org/10.1787/008232ec-en},
  \url{https://www.oecd-ilibrary.org/content/paper/008232ec-en}

\bibitem{9758585}
Ozkaya, I.: Understanding and building trust in software systems. IEEE Software
   \textbf{39}(3), ~3--6 (2022). \doi{10.1109/MS.2021.3134112}

\bibitem{selected11}
van Pelt, R., Jansen, S., Baars, D., Overbeek, S.: Defining blockchain
  governance: A framework for analysis and comparison. Information Systems
  Management  \textbf{38}(1),  21--41 (2021).
  \doi{10.1080/10580530.2020.1720046}

\bibitem{iota}
Popov, S.: The tangle. White paper  \textbf{1}(3), ~30 (2018)

\bibitem{10.1007/3-540-45518-3_18}
Rowstron, A., Druschel, P.: Pastry: Scalable, decentralized object location,
  and routing for large-scale peer-to-peer systems. In: Guerraoui, R. (ed.)
  Middleware 2001. pp. 329--350. Springer Berlin Heidelberg, Berlin, Heidelberg
  (2001)

\bibitem{SHARMA201516}
Sharma, A., Kumar, M., Agarwal, S.: A complete survey on software architectural
  styles and patterns. Procedia Computer Science  \textbf{70},  16--28 (2015),
  proceedings of the 4th International Conference on Eco-friendly Computing and
  Communication Systems

\bibitem{siau2018building}
Siau, K., Wang, W.: Building trust in artificial intelligence, machine
  learning, and robotics. Cutter business technology journal  \textbf{31}(2),
  47--53 (2018)

\bibitem{spectre}
Sompolinsky, Y., Lewenberg, Y., Zohar, A.: Spectre: A fast and scalable
  cryptocurrency protocol. Cryptology ePrint Archive, Paper 2016/1159 (2016),
  \url{https://eprint.iacr.org/2016/1159},
  \url{https://eprint.iacr.org/2016/1159}

\bibitem{phantom}
Sompolinsky, Y., Zohar, A.: Phantom. IACR Cryptology ePrint Archive, Report
  2018/104  (2018)

\bibitem{Spaho2914}
Spaho, E., Sakamoto, S., Barolli, L., Xhafa, F., Ikeda, M.: Trustworthiness in
  p2p: performance behaviour of two fuzzy-based systems for jxta-overlay
  platform. Soft Computing  \textbf{18}(9),  1783--1793 (2014).
  \doi{10.1007/s00500-013-1206-4},
  \url{https://doi.org/10.1007/s00500-013-1206-4}

\bibitem{van2016}
van Steen, M., Tanenbaum, A.S.: A brief introduction to distributed systems.
  Computing  \textbf{98}(10),  967--1009 (2016).
  \doi{10.1007/s00607-016-0508-7},
  \url{https://doi.org/10.1007/s00607-016-0508-7}

\bibitem{TAN2022101625}
Tan, E., Mahula, S., Crompvoets, J.: Blockchain governance in the public
  sector: A conceptual framework for public management. Government Information
  Quarterly  \textbf{39}(1),  101625 (2022).
  \doi{https://doi.org/10.1016/j.giq.2021.101625},
  \url{https://www.sciencedirect.com/science/article/pii/S0740624X21000617}

\bibitem{scheuermann2015iacr}
Tschorsch, F., Scheuermann, B.: {Bitcoin and Beyond: {A} Technical Survey on
  Decentralized Digital Currencies}. IEEE Communications Surveys \& Tutorials
  \textbf{18}(3), ~464 (2016)

\bibitem{van2002distributed}
Van~Steen, M.: Distributed systems principles and paradigms. Network
  \textbf{2}, ~28 (2002)

\bibitem{dao}
Wang, Q., Yu, G., Sai, Y., Sun, C., Nguyen, L.D., Xu, S., Chen, S.: An
  empirical study on snapshot daos (2022). \doi{10.48550/ARXIV.2211.15993},
  \url{https://arxiv.org/abs/2211.15993}

\bibitem{initiatives.weforum.org}
weforum.org: Earning digital trust: Decision-making for trustworthy
  technologies insight report (Nov 2022),
  \url{https://www3.weforum.org/docs/WEF_Earning_Digital_Trust_2022.pdf}

\bibitem{9426788}
Xu, X., Dilum~Bandara, H., Lu, Q., Weber, I., Bass, L., Zhu, L.: A decision
  model for choosing patterns in blockchain-based applications. In: 2021 IEEE
  18th International Conference on Software Architecture (ICSA). pp. 47--57
  (2021). \doi{10.1109/ICSA51549.2021.00013}

\bibitem{connector}
Xu, X., Pautasso, C., Zhu, L., Gramoli, V., Ponomarev, A., Tran, A.B., Chen,
  S.: The blockchain as a software connector. In: 2016 13th Working IEEE/IFIP
  Conference on Software Architecture (WICSA). pp. 182--191 (2016).
  \doi{10.1109/WICSA.2016.21}

\bibitem{bc_pattern}
Xu, X., Pautasso, C., Zhu, L., Lu, Q., Weber, I.: A pattern collection for
  blockchain-based applications. In: Proceedings of the 23rd European
  Conference on Pattern Languages of Programs. EuroPLoP '18, Association for
  Computing Machinery, New York, NY, USA (2018). \doi{10.1145/3282308.3282312},
  \url{https://doi.org/10.1145/3282308.3282312}

\bibitem{7930224}
Xu, X., Weber, I., Staples, M., Zhu, L., Bosch, J., Bass, L., Pautasso, C.,
  Rimba, P.: A taxonomy of blockchain-based systems for architecture design.
  In: 2017 IEEE International Conference on Software Architecture (ICSA). pp.
  243--252 (2017). \doi{10.1109/ICSA.2017.33}

\bibitem{dao2}
Yu, G., Wang, Q., Bi, T., Chen, S., Xu, S.: Leveraging architectural approaches
  in web3 applications--a dao perspective focused. arXiv preprint
  arXiv:2212.05314  (2022)

\bibitem{ironfoge}
Yu, G., Wang, X., Sun, C., Wang, Q., Yu, P., Ni, W., Liu, R.P., Xu, X.:
  Ironforge: An open, secure, fair, decentralized federated learning (2023).
  \doi{10.48550/ARXIV.2301.04006}, \url{https://arxiv.org/abs/2301.04006}

\bibitem{9233457}
Zhang, W., Lu, Q., Yu, Q., Li, Z., Liu, Y., Lo, S.K., Chen, S., Xu, X., Zhu,
  L.: Blockchain-based federated learning for device failure detection in
  industrial iot. IEEE Internet of Things Journal  \textbf{8}(7),  5926--5937
  (2021). \doi{10.1109/JIOT.2020.3032544}

\bibitem{8029379}
Zheng, Z., Xie, S., Dai, H., Chen, X., Wang, H.: An overview of blockchain
  technology: Architecture, consensus, and future trends. In: 2017 IEEE
  International Congress on Big Data (BigData Congress). pp. 557--564 (2017).
  \doi{10.1109/BigDataCongress.2017.85}

\bibitem{Zhu2022}
Zhu, L., Xu, X., Lu, Q., Governatori, G., Whittle, J.: Ai
  and ethics---operationalizing responsible ai pp. 15--33 (2022)

\end{thebibliography}

\end{document}